\documentclass[aps,showpacs,amsmath,amssymb,twocolumn,pra,superscriptaddress,notitlepage]{revtex4-2}

\usepackage{amsmath,bm}
\usepackage[dvipdfmx]{graphicx}
\usepackage{amsmath,amssymb,amsthm,mathrsfs,amsfonts,dsfont}
\usepackage{braket}
\usepackage{color}
\usepackage{graphicx}
\usepackage[linesnumbered, ruled,vlined]{algorithm2e}
\usepackage{braket}
\usepackage{comment}
\usepackage{appendix}
\usepackage{here}
\usepackage{tabularx}
\usepackage{multirow}
\usepackage{mathtools}
\usepackage{qcircuit}
\usepackage{enumerate}
\usepackage{subfigure, epsfig}
\usepackage{mhchem}

\newcolumntype{Y}{>{\centering\arraybackslash}X}

\begin{document} 

\title{Quantum algorithm for a chemical reaction path optimization by using a variational quantum algorithm and a reaction path generation}

\author{Shu Kanno} \email{kanno.s.ac@m.titech.ac.jp} \affiliation{Department of Materials Science and Engineering, Tokyo Institute of Technology, 2-12-1 O-okayama, Meguro-ku, Tokyo 152-8552, Japan}

\begin{abstract} 
The search for new computational tasks of quantum chemistry that can be performed on current quantum computers is important for the development of quantum computing and quantum chemistry. 
Although calculations of chemical reactions have a wide range of applications in quantum chemical calculations, a quantum algorithm for obtaining activation energy $E_a$, which determines the rate of chemical reactions, has not been performed. 
In this study, we propose a quantum algorithm for the chemical reaction path optimization to obtain $E_a$. 
In our algorithm, quantum circuits can be used not only for the energy evaluation using the variational quantum eigensolver (VQE) but also for chemical reaction path generation. 
The chemical reaction path is obtained by encoding the initial reaction path to the circuit, operating parameterized gates, and extracting the path information by measurement.
The nudged elastic band method was used for optimizing a reaction path, and the ground-state calculation for each state on the path was performed by the VQE or the exact diagonalization (ED). 
The proposed algorithm was applied to \ce{H2 + H -> H2 + H} reaction, and we confirmed that $E_a$ was obtained accurately in the case of both the VQE and the ED. 
We also obtained numerical results that the entanglement between the images accelerates the path optimization. 
From these results, we show the feasibility of performing fast and accurate chemical reaction calculations by using quantum algorithms. 
\end{abstract}

\maketitle

\section{Introduction} 
\label{Introduction} 
Chemical reaction calculations have a wide range of applications in the materials developments, such as catalysts~\cite{Reiher2017-jt,Norskov2009-fr} and batteries~\cite{Rice2021-si}.
The chemical reaction speed is determined by the reaction rate constant $k$ which depends on the activation energy $E_a$ in the form of an exponent as $k \propto \mathrm{exp}(E_a/RT)$, where $R$ is the gas constant and $T$ is the temperature. 
Hence, calculating $E_a$ with high accuracy is required for predicting chemical reactions. 

The examples of the reaction path optimization methods for obtaining $E_a$ include the quadratic synchronous transit method~\cite{Govind2003-ni}, intrinsic reaction coordinate method~\cite{Fukui1981-gp}, dimer method~\cite{Henkelman1999-sh}, and nudged elastic band (NEB) method~\cite{Jonsson1998-gi,Henkelman2000-sp,Henkelman2000-zo}. 
The NEB method searches for a minimum energy path (MEP), which has the smallest energy maximum for the reaction path connecting a reactant and a product on a potential energy surface given in reaction coordinates (e.g., bond lengths and angles).
Concretely, the method searches the reaction path on the potential energy surface by generating several intermediate states, called “images”, between the reaction coordinates corresponding to a reactant (initial state) and a product (final state). 
The adjacent images are connected by virtual springs along the reaction path to maintain an equal distance between the images during the optimization, where the optimization of the intermediate images is performed so that the maximum energy of the images becomes smaller.
Here, the ground-state energy on each image is calculated by using an electronic structure calculation method such as the density functional theory (DFT).
$E_a$ is obtained from the optimized reaction path. 

While the NEB method is widely used in chemical reaction path calculations, there are mainly two obstacles to the NEB method for accurately obtaining $E_a$. 
The first obstacle is a lack of accuracy in the ground-state energy calculation. 
The common electronic structure calculation methods such as the DFT may lack the accuracy required for predicting chemical reactions when the electronic correlation of the system is essential to the calculation~\cite{Weymuth2014-le}. 
The second obstacle is a slow convergence of path optimization. Since the images interact with each other in the NEB method, the convergence speed of the optimization depends on the handling procedure of the interactions, and thus we sometimes struggle with slow convergence, especially in a complex chemical reaction path~\cite{Henkelman2000-zo,Asada2018-ix,Galvan2008-at}.

Meanwhile, a quantum computer has attracted considerable attention as it can perform some tasks more efficiently than a classical computer, and the quantum computer has the potential to solve the above obstacles.
For example, the quantum algorithm, called variational quantum eigensolver (VQE)~\cite{Peruzzo2014-kp,Cao2019-qa,McArdle2020-fz}, is expected to solve the first obstacle.
The VQE is an algorithm for obtaining the ground-state energy by repeating the generation of a (variational) wave function by using a parameterized quantum circuit and the update of the parameters on a classical computer. 
In the VQE, the multipartite quantum entanglement between the qubits is expected to enable us to calculate the wave functions that cannot be represented by classical computers efficiently. 
In addition, while current quantum computers, called noisy intermediate-scale quantum (NISQ) devices~\cite{Preskill2018-sc}, are not free from noises in quantum gate operations, the results of the VQE are robust to physical noise since only short depth quantum circuits are used.  
The VQE has been conducted for the calculation of molecules by using NISQ devices~\cite{Peruzzo2014-kp,Shen2017-yt}. 
Furthermore, a quantum computer is expected to solve the second obstacle also since it has been confirmed that the multipartite entanglements in quantum circuits affect the convergence speed of the objective function in optimization tasks~\cite{Abbas2021-ki,Patti2021-jc}.
However, despite these circumstances, a quantum algorithm for obtaining $E_a$ has not been performed.

In this study, we propose a quantum algorithm for chemical reaction path optimization.   
In our algorithm, quantum circuits can be used not only for the ground-state calculation but also for the chemical reaction path generation.
The ground state of each image is calculated by using the VQE (or the exact diagonalization, ED, for comparison). 
We demonstrated our quantum algorithm by applying it to \ce{H2 + H -> H + H2} reaction. 
We found that the proposed algorithm correctly optimizes the reaction path and accurately obtains $E_a$ from the reaction path. 
We also examined the dependencies of the convergence behavior of the path optimization on the entanglers in the path generating quantum circuits and confirmed that the existence of the entanglers accelerates the path optimization, especially when the number of images is large. 
The results show that the proposed quantum algorithm expects to accelerate a chemical reaction path optimization and obtain accurate activation energy compared to the classical algorithms.

\section{Method} \label{Method} We first give an overview of the NEB method and then explain the proposed algorithm for the reaction path optimization. 

\subsection{Overview of the NEB method} 
\label{Sec: Overview of the NEB method} 
We first generate $N_{image}$ images between the reaction coordinate of the reactant and that of the product on the potential energy surface. 
Then, we create a reaction path by connecting adjacent images using virtual springs to obtain the force on each image in the reaction path.
The evaluation value $\bar{F}$ is defined as the average of the norms of the forces on each image,
\begin{equation}     
\begin{aligned}     
\bar{F}=\frac{1}{N_{image}} \sum_{i=1}^{N_{image} }|\mathbf{F}_{i}|,     
\label{Eq: Fbar}
\end{aligned} 
\end{equation} 
where $\mathbf{F}_i$ is the force on the $i$-th image ($i=1,2, \dots, N_{image}$).
$\mathbf{F}_i$ is composed of the spring force $\mathbf{F}_i^S$ and the gradient for the potential energy surface $\boldsymbol{\nabla}E(\mathbf{R}_i)$, 
\begin{equation}     
\begin{aligned} 
\mathbf{F}_{i}=\mathbf{F}_i^S|_{\|}-\boldsymbol{\nabla} E(\mathbf{R}_{i})_{\perp},
\end{aligned} 
\end{equation}
where $\mathbf{R}_i$ is the reaction coordinate in the $i$-th image, and $E(\mathbf{R}_i)$ is the energy of the system specified by $\mathbf{R}_i$. $\mathbf{F}_i^S|_{\|}$ is the tangential component for the path of $\mathbf{F}_i^S$, and $\boldsymbol{\nabla} E(\mathbf{R}_{i})_{\perp}$ is the tangential perpendicular component of $\boldsymbol{\nabla} E(\mathbf{R}_{i})$,
\begin{equation}     
\begin{aligned} 
\left.\mathbf{F}_{i}^{S}\right|_{\|} &=\left(\mathbf{F}_{i}^{S} \cdot \hat{\boldsymbol{\tau}}_{i}\right) \hat{\boldsymbol{\tau}}_{i}\\
&=K\left(\left|\mathbf{R}_{i+1}-\mathbf{R}_{i}\right|-\left|\mathbf{R}_{i}-\mathbf{R}_{i-1}\right|\right) \hat{\boldsymbol{\tau}}_{i}     
\end{aligned} 
\end{equation}
\begin{equation}     
\begin{aligned} 
\boldsymbol{\nabla} E(\mathbf{R}_{i})|_{\perp}=\boldsymbol{\nabla} E(\mathbf{R}_{i})-\boldsymbol{\nabla} E(\mathbf{R}_{i}) \cdot \hat{\boldsymbol{\tau}}_{i},
\label{Eq: Grad E}
\end{aligned} 
\end{equation}
where ${\hat{\boldsymbol{\tau}}}_i$ is the tangent vector, and $K$ is the spring constant.
${\hat{\boldsymbol{\tau}}}_i$ is defined as 
\begin{equation}     
\begin{aligned}     
    \hat{\boldsymbol{\tau}}_i = \begin{cases} \hat{\boldsymbol{\tau}}_i^+ & \mathrm{if}\ E(\mathbf{R}_{i+1})>E(\mathbf{R}_{i})>E(\mathbf{R}_{i-1}) \\\hat{\boldsymbol{\tau}}_i^- & \mathrm{if}\ E(\mathbf{R}_{i+1})<E(\mathbf{R}_i)<E(\mathbf{R}_{i-1}) \end{cases},  
    \label{eq: Tangent 1}
    \end{aligned} 
\end{equation} 
else if $E(\textbf{R}_{i+1})<E(\textbf{R}_{i})>E(\textbf{R}_{i-1})$ or $E(\textbf{R}_{i+1})>E(\textbf{R}_{i})<E(\textbf{R}_{i-1})$
\begin{equation} 
\begin{aligned} 
&\hat{\boldsymbol{\tau}}_i = \begin{cases} \hat{\boldsymbol{\tau}}_i^+\Delta E_i^{\mathrm{max}}+\hat{\boldsymbol{\tau}}_i^- \Delta E_i^{\mathrm{min}} & \mathrm{if}\ E(\mathbf{R}_{i+1})>E(\mathbf{R}_{i-1}) \\\hat{\boldsymbol{\tau}}_i^+\Delta E_i^{\mathrm{min}}+\hat{\boldsymbol{\tau}}_i^- \Delta E_i^{\mathrm{max}} & \mathrm{if}\ E(\mathbf{R}_{i+1})<E(\mathbf{R}_{i-1}) \end{cases},
\label{eq: Tangent 2}
\end{aligned}     
\end{equation} 
where $\hat{\boldsymbol{\tau}}_i^+ =\mathbf{R}_{i+1}-\mathbf{R}_{i} $, $\hat{\boldsymbol{\tau}}_i^- = \mathbf{R}_{i}-\mathbf{R}_{i-1}$, $\Delta E_i^{\mathrm{max}}=\mathrm{max}(|E(\mathbf{R}_{i+1})-E(\mathbf{R}_{i})|,|E(\mathbf{R}_{i-1})-E(\mathbf{R}_{i})|)$, and $\Delta E_i^{\mathrm{min}}=\mathrm{min}(|E(\mathbf{R}_{i+1})-E(\mathbf{R}_{i})|,|E(\mathbf{R}_{i-1})-E(\mathbf{R}_{i})|)$.
${\hat{\boldsymbol{\tau}}}_i$ needs to be normalized. 
Note that the tangential perpendicular component of $\mathbf{F}_i^S$, $\mathbf{F}_i^S|_{\perp}$, and the tangential component of $\boldsymbol{\nabla} E(\mathbf{R}_{i})$, $\boldsymbol{\nabla} E(\mathbf{R}_{i})_{\|}$,  are not used in the NEB method to prevent the shifting of the image from the MEP and to keep the distance between the images equal, respectively~\cite{Jonsson1998-gi}. 
In addition, ${\hat{\boldsymbol{\tau}}}_i$ in Eqs.~(\ref{eq: Tangent 1}) and (\ref{eq: Tangent 2}) are defined for preventing the path from oscillating during the convergence process~\cite{Henkelman2000-sp}.
$N_{image}$ was set to be three, five, or seven, and $K$ was set to be 0.1 Hartree/Å$\mathrm{^2}$.

\subsection{Optimization of the reaction path by using the path generation through quantum circuits}
\label{Sec: Optimization of reaction path}
\begin{figure}[!ht]  
\includegraphics[width=1\columnwidth]{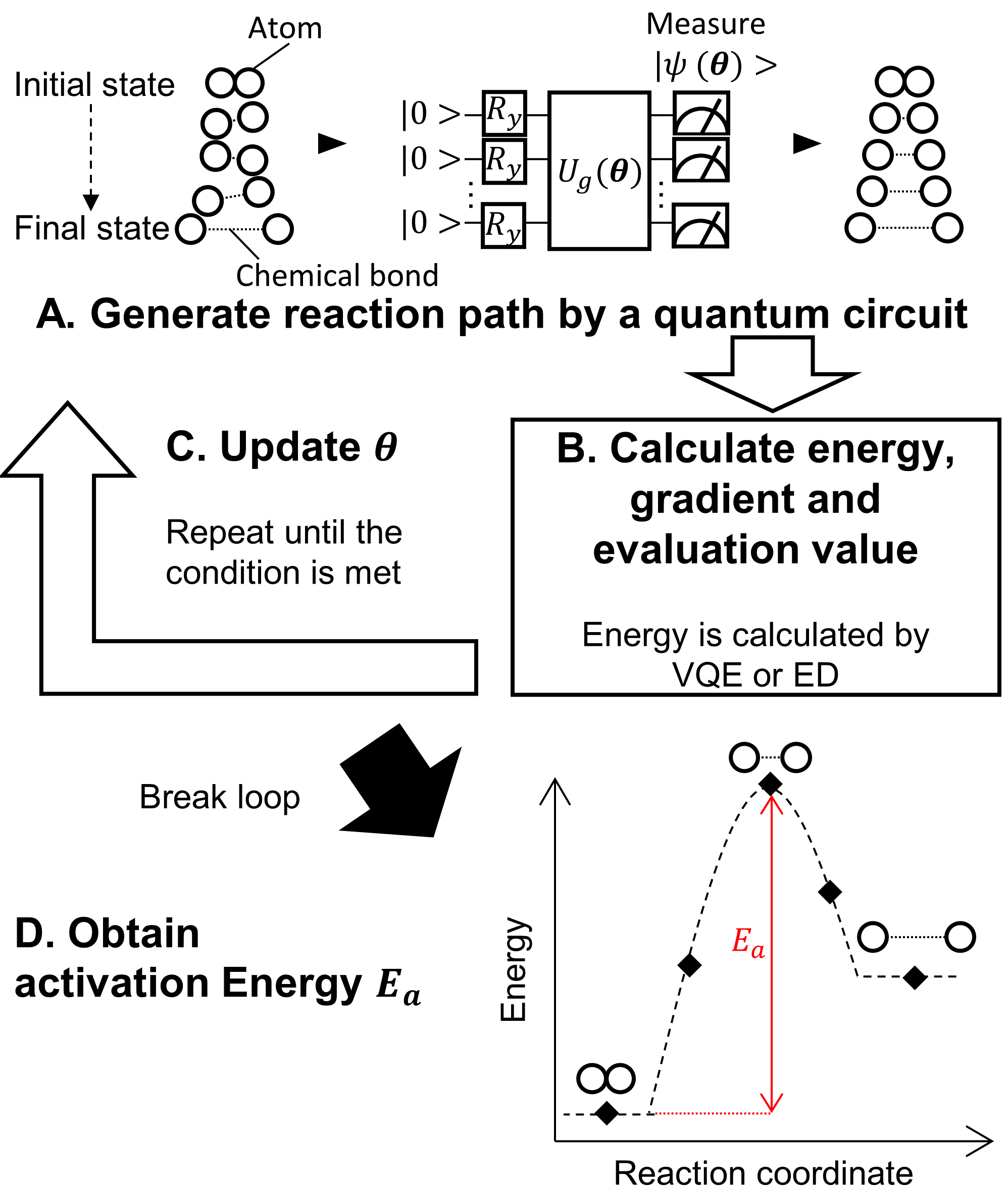} 
\caption{The calculation flow of optimizing a reaction path and obtaining the activation energy $E_a$ by using the path generation though quantum circuits.}  
\label{fig: Overview} 
\end{figure}

Next, we explain the detail of the proposed algorithm. 
Figure \ref{fig: Overview} shows a calculation flow of the proposed quantum algorithm for obtaining $E_a$. 
The outline of the calculation is shown in the following four steps, and the capital letters in each step correspond to the letters in Fig.~\ref{fig: Overview}.
\begin{enumerate}[Step A.]
    \item Generate the reaction path (the geometries) using a parameterized quantum circuit ($\boldsymbol{\theta}$ is a set of the parameters).
    \item Calculate the ground-state energy $E(\mathbf{R}_{i})$ and the gradient $\boldsymbol{\nabla} E(\mathbf{R}_{i})$ and obtain the evaluation value $\bar{F}$.
    \item Calculate the gradient for $\boldsymbol{\theta}$ of $\bar{F}$, $\boldsymbol{\nabla}_{\boldsymbol{\theta}}\bar{F}$, and update  $\boldsymbol{\theta}$.
    \item Repeat steps A to C until the termination condition is satisfied. 
    After the termination, $E_a$ is obtained from the optimized reaction path.
\end{enumerate}

The details for each step are as follows.

\begin{center}
\emph{Step A. Reaction path generation method by using quantum circuits}
\end{center}

\begin{figure}[!ht]  
\includegraphics[width=1\columnwidth]{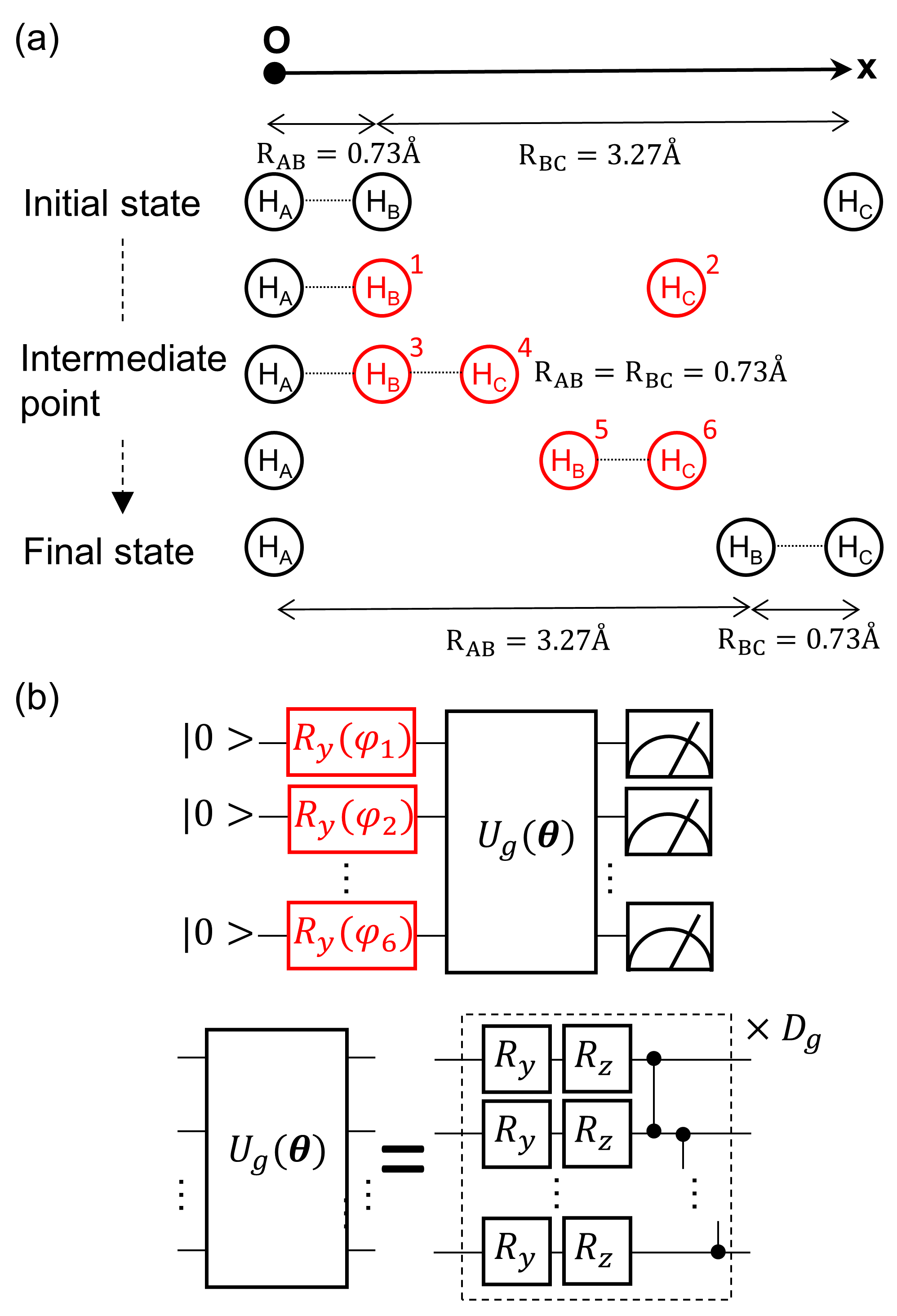}  
\caption{Initial reaction path for the H\textsubscript{2} + H → H + H\textsubscript{2} reaction and the quantum circuit used for generating the path. 
We show the example of $N_{image} = 3$.
(a) Depiction of the reaction path. 
Atoms with fixed (unfixed) positions in the reaction path optimization calculations are shown in black (red). 
(b) Details of the circuit.
The upper panel show whole circuit implementation.
The lower panel show details of $U_{g}(\boldsymbol{\theta})$. 
The block surrounded by the dashed line was repeated $D_{g}$ times.}  
\label{fig: Initial reaction path} 
\end{figure}

Figure~\ref{fig: Initial reaction path}(a) show the depiction of the reaction path, and we assume the reaction of H\textsubscript{2} + H → H + H\textsubscript{2} with one-dimensional system and $N_{image}=3$.
Each row corresponds to any of the initial state (IS), the image, or the final state (FS).
$(N_{image} - 1)/2 = 1$ image is interpolated each between the IS and the intermediate point (IMP) and between the IMP and the FS. 
In each row, the three hydrogen atoms are named H\textsubscript{A}, H\textsubscript{B}, and H\textsubscript{C} from left to right, and the position of H\textsubscript{A} is defined as the origin. 
The distance between H\textsubscript{A} and H\textsubscript{B} (H\textsubscript{B} and H\textsubscript{C}) is described as $\text{R}_{\text{AB}}$ ($\text{R}_{\text{BC}}$). 
We assumed $\text{R}_{\text{AB}}$ and $\text{R}_{\text{BC}}$ as the reaction coordinates on the potential energy surface.

First, we explain encoding the single image into a circuit. 
The distances in the Cartesian coordinates are divided by the reference length $\text{R}_{\text{ref}}\left( \geq \text{R}_{\text{AB}} + \text{R}_{\text{BC}} \right)$ to convert them to the fractional coordinates of $[0, 1]$. 
Specifically, the coordinates of H\textsubscript{A}, H\textsubscript{B}, and H\textsubscript{C} are converted from $0$, $\text{R}_{\text{AB}}$, and $\text{R}_{\text{AB}} + \text{R}_{\text{BC}}$ to $0$, $\text{R}_{\text{AB}}/\text{R}_{\text{ref}}$, and $( \text{R}_{\text{AB}} + \text{R}_{\text{BC}} )/\text{R}_{\text{ref}}$, respectively. 
Here, $\text{R}_{\text{ref}}$ was set to be 6.00 Å. 
Then, the fractional coordinate is encoded to the wave function as $R_y(2\arccos(r))\ket{0}$, where $R_{y}(\varphi)$ is $R_y$ gate with a rotation angle $\varphi$, and $r$ is a fractional coordinate. 
Specifically, the corresponding states are respectively $R_y(0)\ket{0} = \ket{0}$, 
$R_y(2\arccos( \sqrt{\text{R}_{\text{AB}}/\text{R}_{\text{ref}}}))\ket{0} = \sqrt{\text{R}_{\text{AB}}/\text{R}_{\text{ref}}}\ket{0} + \sqrt{1 - \text{R}_{\text{AB}}/\text{R}_{\text{ref}}}\ket{1}$, 
and $R_y(2\arccos( \sqrt{( \text{R}_{\text{AB}} + \text{R}_{\text{BC}})/\text{R}_{\text{ref}}} ))\ket{0} = \sqrt{( \text{R}_{\text{AB}} + \text{R}_{\text{BC}})/\text{R}_{\text{ref}}}\ket{0} + \sqrt{1 - ( \text{R}_{\text{AB}} + \text{R}_{\text{BC}})/\text{R}_{\text{ref}}}\ket{1}$. 
We can encode the single image into a circuit by arranging the wave functions, i.e., $\ket{\Psi_i} = R_y(0)\ket{0} \otimes R_y(2\arccos( \sqrt{\text{R}_{\text{AB}}/\text{R}_{\text{ref}}}))\ket{0} \otimes R_y(2\arccos( \sqrt{( \text{R}_{\text{AB}} + \text{R}_{\text{BC}})/\text{R}_{\text{ref}}} ))\ket{0}$.

We can also encode the path by arranging the image-encoded states $\ket{\Psi_i}$ on a circuit, i.e., $\bigotimes_{i=1}^{N_{image}}\ket{\Psi_i}$. 
However, it is not necessary to encode the geometrical information of the atoms of which the coordinates are fixed in the NEB calculation.
As shown in Figure~\ref{fig: Initial reaction path}(a), the IS, the FS, and the positions of H\textsubscript{A} are fixed (black in the Figure).
Since 6 unfixed atoms (red in the Figure) are present among a total of 15 atoms in the path including the IS and the FS, the number of qubits for obtaining the path-encoded state $\ket{\psi_{enc}}$ can be reduced from 15 to 6.
We note that the number of qubits required for the path generation of an $N_{dim}$-dimensional system is $N_{dim}N_{UFA} \sim O( N_{dim}N_{atom}N_{image})$, where $N_{UFA}$ is the number of the unfixed atoms, and $N_{atom}$ is the number of atoms in a image.

The circuit implementation for generating the path is shown in Fig.~\ref{fig: Initial reaction path}(b).
The number assigned to each atom in red in Fig.~\ref{fig: Initial reaction path}(a) corresponds to the qubit index $m$ $(m = 1,\ 2, \dots, 6)$ in Fig.~\ref{fig: Initial reaction path}(b). 
We create the wave functions of the entangled geometrical information, $\ket{\psi(\boldsymbol{\theta})}$, by operating unitary gates with parameters $U_{g}(\boldsymbol{\theta})$ on $\ket{\psi_{enc}}$, i.e., $\ket{\psi(\boldsymbol{\theta})} = U_{g}(\boldsymbol{\theta})\ket{\psi_{enc}}$. 
The quantum entanglement expects to allow to take a different behavior of path optimization convergence compared with the classical algorithms (see results in Sec.~\ref{sec: Convergence of the evaluation value} and Appendix~\ref{sec: Futher study of convergence}).
The parameterized unitary gates $U_{g}(\boldsymbol{\theta})$ were implemented based on Hardware efficient ansatz~\cite{Kandala2017-lh}. 
All initial $\boldsymbol{\theta}$ were set to zero, and the depth of the path generating circuit $D_{g}$ (lower panel of Fig.~\ref{fig: Initial reaction path}(b)) was set to be two. 

Then, the path (the geometrical information) are extracted from the wave function. 
We make the probability of measuring $m$-th qubit of the circuit, $P_m$, correspond to the fractional coordinate of the $m$-th atom,
\begin{equation}
    \begin{aligned}
    P_m = |\langle 0|_{m}|\psi(\boldsymbol{\theta})\rangle|^{2},
    \label{Eq: Frac}
    \end{aligned}
\end{equation}
where $\bra{0}_{m}$ represents $\bra{0}$ acting on the $m$-th qubit. 
We mention that in the definition in Eq.~(\ref{Eq: Frac}), $P_m$ corresponds to the initial fractional coordinate of the $m$-th atom when $U_g(\boldsymbol{\theta}) = I$, where $I$ is the identity operator.
All quantum circuit simulations in this work were classically implemented by using Blueqat package~\cite{Mochizuki2019-oa}.

We note two points about quantum advantages in the path generation method. Firstly, the expressible reaction path by the circuit does not change with or without the entanglement (i.e., the $\text{CZ}$ gate) since all the geometries can be represented without $U_g(\boldsymbol{\theta})$ by changing the rotation angle of $R_{y}$ gate in $\ket{\psi_{enc}}$. 
Our expectation for quantum entanglement in the path generating circuit is to improve the expressive power of the reaction path update, not of the reaction path itself. 
Secondly, in this study, we do not discuss the quantum advantage of encoding the path information to quantum circuits in this study although the efficient encoding of classical data to a quantum circuit is itself a remarkable research topic~\cite{Schuld2018-vy}.

\begin{center}
\emph{Step B. Energy, gradient, and evaluation value calculation}
\end{center}

The values of $E(\mathbf{R}_{i})$ and $\boldsymbol{\nabla}E(\mathbf{R}_{i})$ are obtained by calculating the ground-state energy at the geometries corresponding to $\mathbf{R}_{i}$ and around $\mathbf{R}_{i}$. 
$\bar{F}$ is obtained by substituting the values $\mathbf{R}_{i}$, $E(\mathbf{R}_{i})$, and $\boldsymbol{\nabla}E(\mathbf{R}_{i})$ into Eqs.~(\ref{Eq: Fbar})-(\ref{eq: Tangent 2}).

\begin{figure}[!t]  
\includegraphics[width=1\columnwidth]{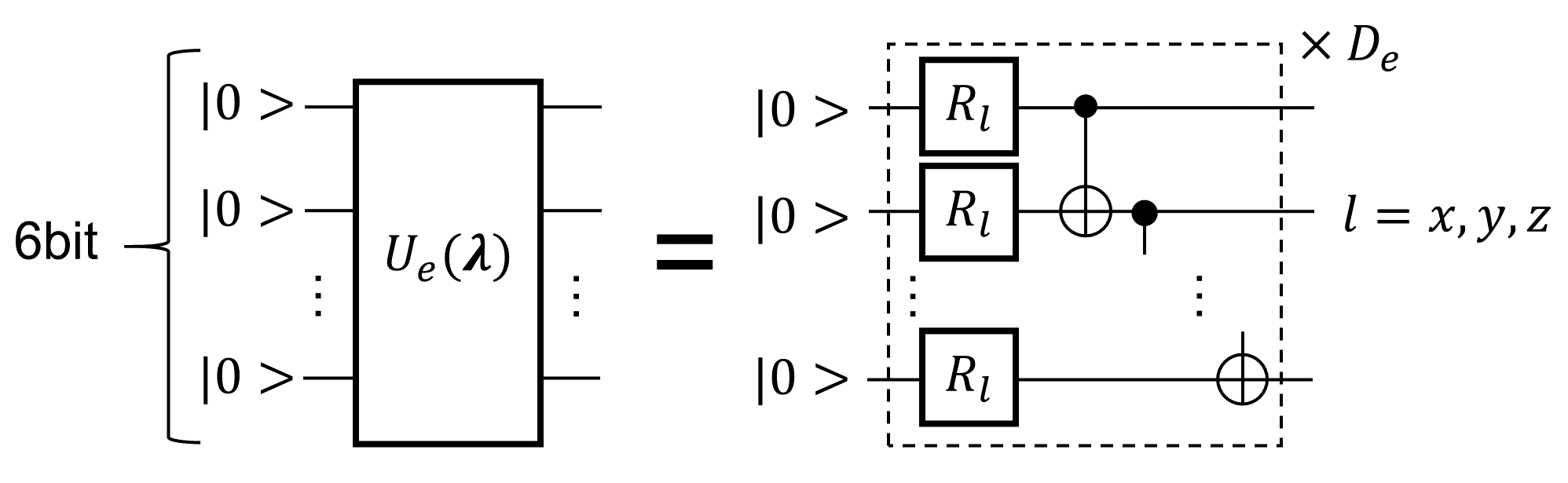}  
\caption{Quantum circuit used in the VQE. 
The number of qubits (= spin orbitals) were six. 
The unitary gate $U_{e}(\boldsymbol{\lambda})$ (left panel) was composed of blocks of $R_{l}$ $(l = x,y,z )$ and $\text{CNOT}$ gates within the dashed lines (right panel), and the block surrounded by the dashed line was repeated $D_{e}$ times.}
\label{fig: Quantum circuit} 
\end{figure}

The computational details are as follows.
The Hamiltonians used in the ground-state calculations were created using the OpenFermion package~\cite{McClean2020-zp}. 
The molecular wave functions of the converged Hartree-Fock calculations by STO-3G were used, and the number of spin orbitals was six, i.e., two per single hydrogen atom. 
The ground-state calculation was performed by using the VQE and the ED. 
Figure~\ref{fig: Quantum circuit} shows the quantum circuit used in the ground-state calculation of the VQE. 
The optimization in the VQE was based on a gradient-free method, Rotoselect~\cite{Ostaszewski2021-yk}. 
In the Rotoselect, not only the parameters $\boldsymbol{\lambda}$ but also the types of the rotation gates in the circuit are optimized.
The initial parameters of $\boldsymbol{\lambda}$ were set to be random numbers of $[0.0, 0.1)$, and the initial rotation gates were set to be $R_{x}$. 
The depth of the quantum circuit $D_{e}$ was set to be five. 
The energy evaluation of the VQE was repeated until the energy difference for the VQE iteration became less than $10^{-4}$ Hartree. 
$\boldsymbol{\nabla}E(\mathbf{R}_{i})$ was obtained by using the central difference method, and the difference value for each reaction coordinate was set to 0.1 Å. 

\begin{center}
\emph{Step C. Calculation of the parameter gradient of the evaluation value and parameter update}
\end{center}

By performing steps A and B, we calculate   $\boldsymbol{\nabla}_{\boldsymbol{\theta}}\bar{F}$ and update $\boldsymbol{\theta}$. $\boldsymbol{\nabla}_{\boldsymbol{\theta}}\bar{F}$ was obtained by using the central difference method, and the parameter difference value was set to be 0.001.
$\boldsymbol{\theta}$ was updated using Adam~\cite{Kingma2014-og} with a learning rate of 0.01.

\begin{center}
\emph{Step D. Activation energy evaluation}
\end{center}

We set the termination condition as 100 iterations. 
In the optimized path, $E_{a}$ is defined as the difference between the highest energy among all images and the energy of the IS.

\section{Results}
\label{sec: Results}

\begin{figure*}[!ht]  
\includegraphics[width=0.8\textwidth]{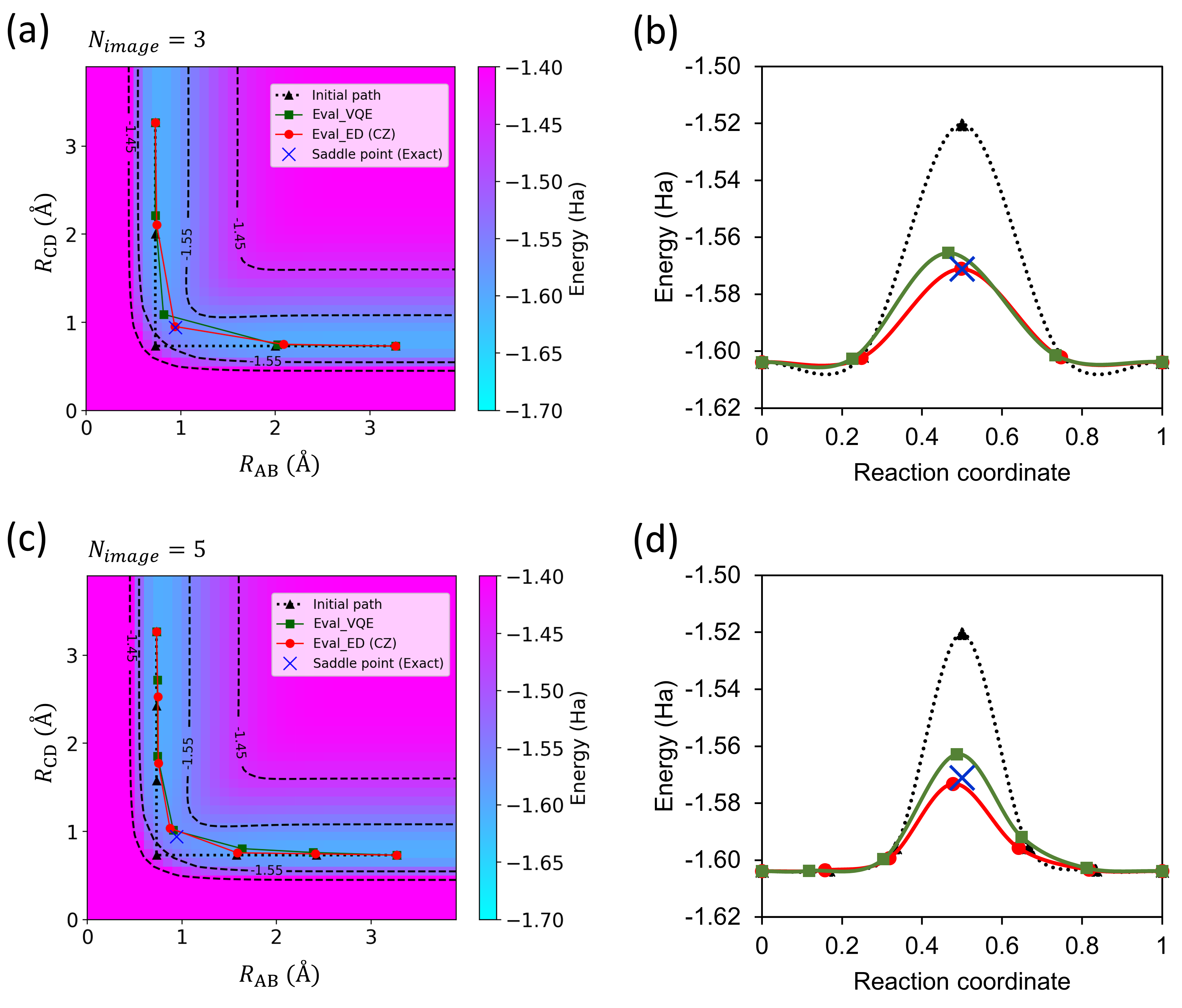}  
\caption{Results of reaction path optimization. 
(a) and (b) [(c) and (d)] are the results for $N_{image} = 3$ [$5$]. 
(a) and (c) are reaction paths on the potential surface.
Each mark corresponds to any of the IS, an image, the FS, and a saddle point state.
The reaction proceeds from the IS at the upper left to the FS at the lower right. 
(b) and (d) show the energies on each reaction path. 
The types of marks and lines in (b) and (d) correspond to those in (a) and (c), respectively.
The reaction proceeds from the IS on the left to the FS on the right. 
The length of each reaction path was normalized to compare the paths of different lengths.}  
\label{fig: Results of reaction} 
\end{figure*}

\begin{table*}[!ht] 
\begin{center} 
\caption{List of obtained values for $N_{image} = 3$ and $5$ after the optimization termination. 
When the values of $N_{image} = 3$ and $5$ were different, the result for 5 images was written in parentheses.} 
\label{tbl: List of Fbar}    
\begin{tabular}{c c c c c c}     3 Images (5 Images) & $\bar{F}$ (Ha/Å) & $\text{R}_{\text{AB}}$ (Å) & $\text{R}_{\text{BC}}$ (Å) & $E_{a}$ (mHa) & $|\Delta_{saddle}|$ (mHa) \\ \hline     
Initial path & 0.17 (0.12) & 0.73 & 0.73 & 83 & 50 \\  Eval\_VQE & 0.02 (0.03) & 0.82 (0.91) & 1.09 (1.01) & 38 (41) & 5 (8) \\  
Eval\_ED & 0.00 (0.00) & 0.94 (0.87) & 0.95 (1.04) & 33 (31) & 0 (2) \\  
Saddle Point & - & 0.94 & 0.94 & 33 & - \\     
\end{tabular} 
\end{center} 
\end{table*}

\begin{figure*}[!ht]  
\includegraphics[width=0.8\textwidth]{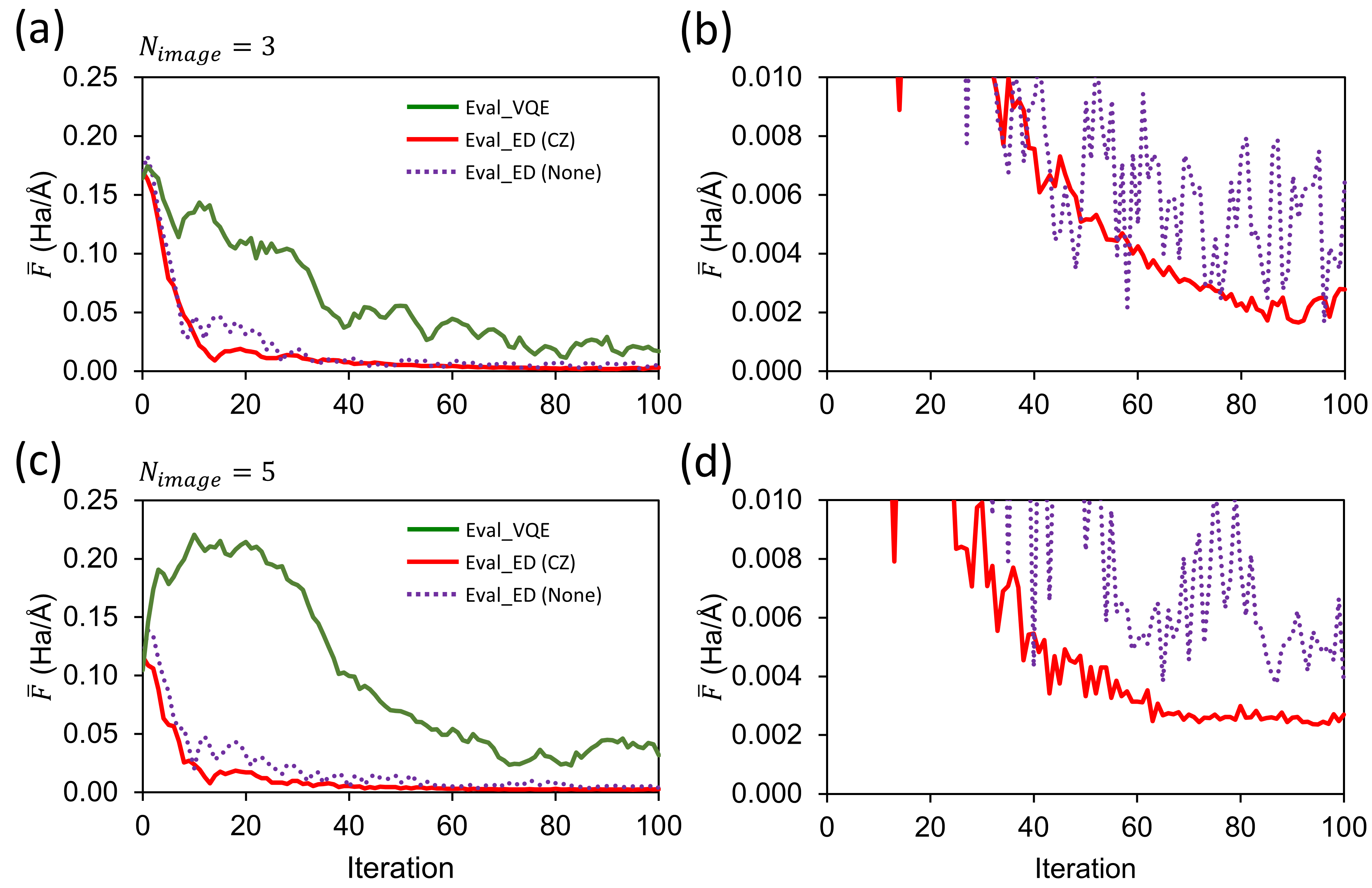}  
\caption{The evaluation value for each iteration. 
(a) and (b) [(c) and (d)] are the results for $N_{image} = 3$ [$5$]. 
(b) and (d) are enlarged views of the vertical axes in (a) and (c), respectively. 
}  
\label{fig: Evaluation value} 
\end{figure*}

\subsection{Numerical results of the proposed algorithm}
\label{sec: Execution results}
Figure~\ref{fig: Results of reaction}(a) and (c) shows the results of the optimized reaction paths of $N_{images} =$ 3 and 5, respectively, on the potential energy surface.
In both Figures, the paths obtained by the NEB method by using the VQE [Eval\_VQE, green line in the Figure] and the ED [Eval\_ED (CZ), red line] become closer to the saddle point [Saddle point (Exact), blue cross mark] and smoother than initial reaction path [Initial path, black dotted line]. 
Furthermore, as shown in Figs.~\ref{fig: Results of reaction}(b) and (d), which are the energies on the reaction paths corresponding to Figs.~\ref{fig: Results of reaction}(a) and (c), respectively, the maximum energies in the reaction paths for both the VQE and the ED are closer to the saddle point energy than that in the initial reaction path.
Thus, we found that the reaction path is correctly optimized by the path generation method using quantum circuits. 
In addition, the VQE results show that quantum computers are useful for optimizing the reaction path since $E_a$ is accurately obtained even when the quantum circuit is used for both the path generation and ground-state calculation.

We mention the effect of the convergence condition of the VQE. 
Table~\ref{tbl: List of Fbar} shows the obtained values, where $|\Delta_{saddle}|$ is an absolute value of the difference between the exact activation energy and the calculated one. 
Although the values of $|\Delta_{saddle}|$ in both the VQE and the ED were sufficiently small compared to that in the initial reaction path, the values in the VQE (5 and 8 mHartree in $N_{image}$ = 3 and 5, respectively) were larger than the chemical accuracy (about 1.6 mHartree). 
However, the accuracy is improved by tightening the convergence condition in the VQE. 
We performed the VQE in the geometry on the saddle point and found that the absolute value of the difference from the exact value decreased from 4.5 mHartree to 0.031 mHartree when the convergence condition was changed from the $10^{-4}$ Hartree (default value) to $10^{-6}$ Hartree. 
Hence, tighter computational conditions than the default may need to be imposed in practical applications of the proposed method.
Note that we choose the default value in order to save the calculation time since multiple numbers of ground-state calculations were needed for the NEB calculation. 
For example, the numbers of performed ground-state calculations per NEB iteration in $N_{image} = 3$ and $5$ were 1225 and 2835, respectively. 
In general, the number of calculations is $O( N_{image}^{2}D_{g}N_{atom}^{2}N_{dim}^{2})$, where $O( N_{image}N_{atom}N_{dim})$ and $O( N_{image}D_{g}N_{atom}N_{dim})$ come from calculating $\bar{F}$ and updating $\boldsymbol{\theta}$, respectively. 

\subsection{Convergence of the evaluation value} 
\label{sec: Convergence of the evaluation value}
Figures~\ref{fig: Evaluation value}(a) and (c) show $\bar{F}$ for each iteration in $N_{image}$ = 3 and 5, respectively. 
In both cases, $\bar{F}$ tends to converge as the number of iterations increases. 
The use of the ED for the ground-state calculation (red line) converged faster than that of the VQE (green line), indicating that the accuracy of the ground-state calculation affects the convergence speed.

Finally, we discuss the effect of the entanglers ($\text{CZ}$ gates) on the convergence. 
The purple dotted lines in Fig.~\ref{fig: Evaluation value} show the convergence when excluding $\text{CZ}$ gate from $U_{g}(\boldsymbol{\theta})$ in Fig.~\ref{fig: Initial reaction path}(b), where the ground-state calculation is performed on the ED.
Figures~\ref{fig: Evaluation value}(b) and (d) are enlarged vertical axes of Figs.~\ref{fig: Evaluation value}(a) and (c), respectively.
We found that the value with $\text{CZ}$ gates tends to be lower than that without $\text{CZ}$ gates after about 50 iterations. 
In addition, from the average results of 10 initial reaction paths (in $N_{image}=3, 5,$ and $7$) generated by slightly changing the atomic positions from Fig.~\ref{fig: Initial reaction path}(a), we confirmed that the trend for CZ gates become clearer as $N_{image}$ increases (see Appendix~\ref{sec: Futher study of convergence}).
Therefore, the entanglement between images would accelerate the convergence of reaction path optimization.

\section{Conclusion}
\label{conclusion}

In this study, we proposed a quantum algorithm for chemical reaction path optimization. 
In our algorithm, we can use the quantum circuit at two steps in the calculation flow: one is the calculation of the ground-state energy, and the other is the generation of the reaction path. 
In the path generating method, the coordinates of each atom in each image of the initial reaction path are encoded to a quantum circuit through the rotation angle of $R_{y}$ gate. 
The encoded state is entangled by using one- and two-qubit gates, and the geometrical information is extracted from the measurement value. 
The nudged elastic band (NEB) method was used for the reaction path optimization. 

We applied the algorithm to optimize the reaction path for the H\textsubscript{2}+H → H + H\textsubscript{2} reaction, where the variational quantum eigensolver (VQE) or the exact diagonalization (ED) were used in the ground-state calculation. 
We found that the reaction paths were accurately optimized, and the activation energy that is close to the saddle point energy was obtained in both the VQE and the ED cases. 
In addition, we confirmed that the entanglement of quantum circuits leads to faster convergence of the NEB method, especially when the number of images is large. 
The results show that the proposed quantum algorithm paves the way to more accurate and faster chemical reaction path optimization.

\section{Acknowledgement} The author would like to thank Tomofumi Tada for useful discussion about the study and for providing computational resources, Takuya Nakao for providing information about the NEB method and for discussing the NEB method in the manuscript, Suguru Endo for discussing the VQE in the manuscript, Asahi Chikaoka for providing information about Rotoselect, Taishi Mizuguchi for exchanging ideas about quantum computing, and Blueqat Co., Ltd.~for various information on quantum computers and quantum simulators. 
\emph{Note added} During the update of this manuscript, several studies of the quantum algorithm for chemical reaction were reported~\cite{Parrish2021-mt,Omiya2022-sq,Yalouz2022-qq}.

\bibliographystyle{apsrev4-1} 
\bibliography{bib} 

\appendix 
\section{Contribution of the entanglers to the convergence speed}
\label{sec: Futher study of convergence}
\begin{figure*}[!ht]  
\includegraphics[width=0.8\textwidth]{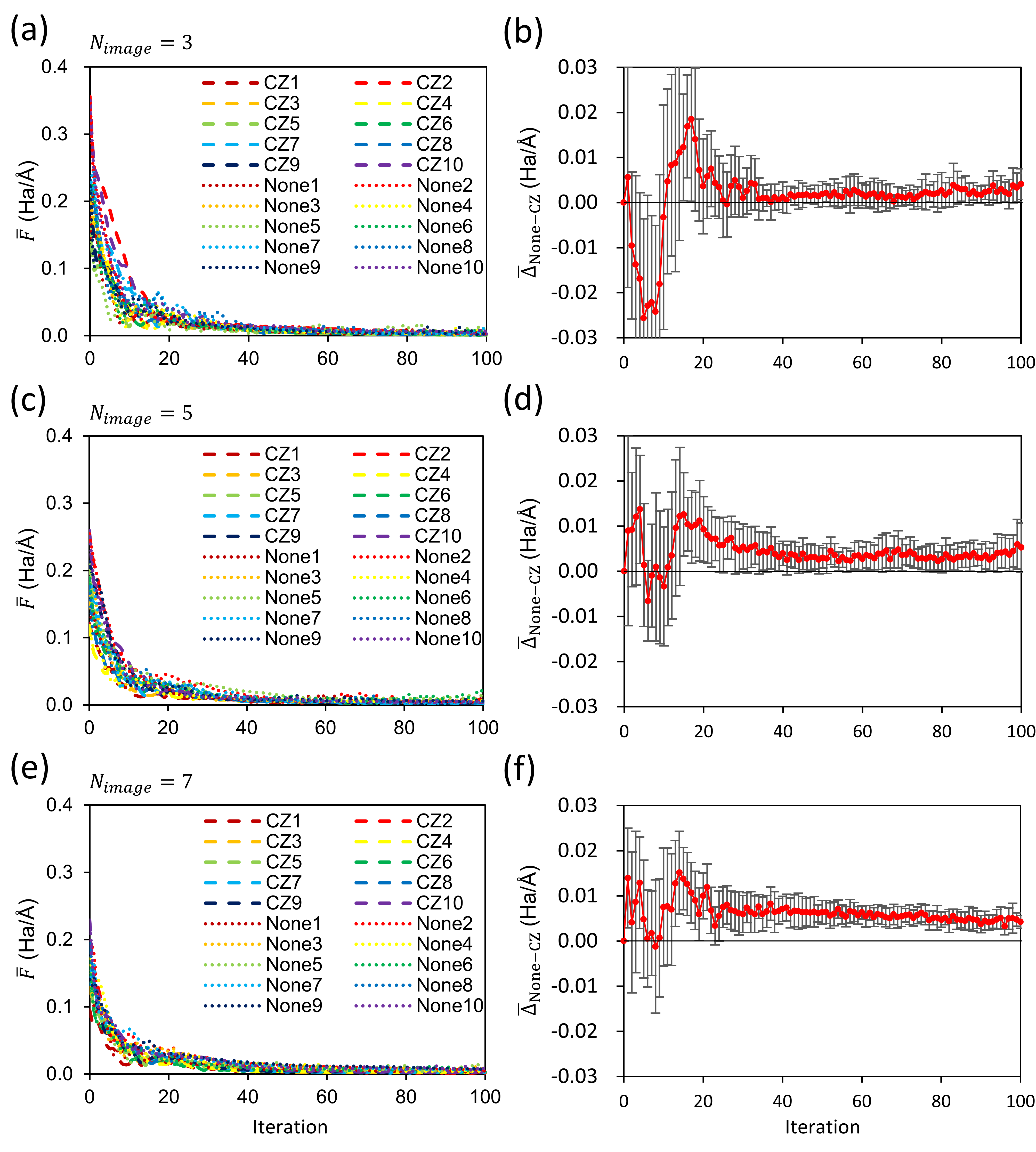}  
\caption{The evaluation value for the 10 initial paths with and without $\text{CZ}$ gates (left) and the average difference of them (right) for each iteration. 
(a) and (b), (c) and (d), and (e) and (f) are the results of$\ N_{image} = 3,\ 5,$ and $7$, respectively. }  
\label{fig: 10 Evaluation} 
\end{figure*}

We examined the behavior of the path optimization convergence with and without $\text{CZ}$ entanglers on multiple initial paths. 
We created 10 initial paths by adding $[-0.1, 0.1)$ random numbers to the values of reaction coordinates $\text{R}_{\text{AB}}$ and $\text{R}_{\text{BC}}$ for each image in the initial path in Fig.~\ref{fig: Initial reaction path}(a).
In the whole calculation, the same 10 random seeds were shared, and the ground-state calculation was performed by the ED.
Figures~\ref{fig: 10 Evaluation}(a), (c) and (e) show the results of path optimizations with and without $\text{CZ}$ gates for the 10 initial paths at $N_{image}$ of $3,\ 5,$ and $7$, respectively. 
Error bars denote the standard deviations. 
While $\bar{F}$ depends on the initial value to some extent, the values tend to converge in any path as the iteration increases.
Figures~\ref{fig: 10 Evaluation}(b), (d), and (f) show the average of the differences of $\bar{F}$ between the values with and without $\text{CZ}$ gate, $\mathrm{\bar{\Delta}_{None\mathchar`-CZ}}$, in Figs~\ref{fig: 10 Evaluation}(a), (c) and (d), respectively. 
In all $N_{image}$, $\mathrm{\bar{\Delta}_{None\mathchar`-CZ}}$ at the iteration $\gtrsim 50$ are positive, that is, the convergence is accelerated by the $\text{CZ}$ gate. 
In addition, the acceleration of the convergence by using the $\text{CZ}$ gates becomes clearer as $N_{image}$ increases.
Especially, in Fig.~\ref{fig: Evaluation value}(f) ($N_{image}=7$), the values at the iteration $\gtrsim50$ are positive even with including the error bar.
The average value at 100 iterations for $N_{image}=7$ is 0.0042 Hartree/Å$\  \approx \ $0.11 eV/Å, which is small but not negligible due to its larger or the same order of magnitude as the typical NEB convergence criteria ($10^{-2}\sim10^{-1}$ eV/Å)~\cite{Henkelman2000-sp,Henkelman2000-zo}.
If the chemical reaction is complex, the path would be longer and the number of required images would be larger.
Therefore, our path generation method has the potential to speed up the chemical reaction path optimization by using quantum entanglements.

\end{document}